\documentclass[11pt]{article}

\usepackage[margin=1in]{geometry}
\usepackage[utf8]{inputenc}
\usepackage[T1]{fontenc}
\usepackage{lmodern}
\usepackage{microtype}
\usepackage{amsmath,amssymb,amsthm}
\usepackage{graphicx}
\usepackage{booktabs}
\usepackage{multirow}
\usepackage{subcaption}
\usepackage{xcolor}
\usepackage{url}
\usepackage[numbers,sort&compress]{natbib}
\usepackage[colorlinks=true,linkcolor=blue,citecolor=blue,urlcolor=black]{hyperref}

\newtheorem{theorem}{Theorem}

\newtheorem{assumption}{Assumption}

\title{\bf Efficient First-Order Methods for Estimating Generalized Additive Index Models}

\author{
Ziyu Peng \quad Linglingzhi Zhu \quad Yao Xie\\[1ex]
H. Milton Stewart School of Industrial and Systems Engineering\\
Georgia Institute of Technology
}

\date{\today}

\begin{document}

\maketitle

\begin{abstract}
Generalized additive index models (GAIMs) offer a flexible semiparametric framework for capturing complex data relationships, balancing the interpretability of parametric models with the flexibility of nonparametric approaches. However, classical stage-wise estimation procedures for GAIMs suffer from computational inefficiencies due to their sequential nature and reliance on nonparametric smoothing. To overcome these drawbacks, we propose efficient, simultaneous estimation algorithms for GAIMs. By leveraging basis expansion, we cast the semiparametric estimation task as a finite-dimensional optimization problem solvable by first-order methods such as gradient descent (GD). Furthermore, we introduce a variational inequality (VI) estimation algorithm, extending the VI framework from generalized linear models to GAIMs. We provide a unified convergence result to a stationary point for both algorithms. Numerical experiments highlight the computational and statistical advantages of our methods over classical stage-wise procedures, and reveal the potential benefits of the VI-based approach over GD for non-canonical link functions.
\end{abstract}

\section{Introduction}

Capturing the relationship between a response variable and a set of predictor variables is a fundamental task in statistics and machine learning. Generalized linear models (GLMs) \cite{nelder1972generalized,mccullagh1989generalized} serve as a widely-used and well-studied tool for modeling such relationships. However, parametric approaches like GLMs may fall short in representation power when the dependency is complex. Conversely, nonparametric regression \cite{hardle1990applied} offers great modeling flexibility, but suffers from the curse of dimensionality both statistically and computationally. To strike a balance between these two paradigms, generalized additive models (GAMs) \cite{stone1985additive,hastie1986generalized,wood2017generalized,wood2025generalized} replace the linear predictor in GLMs with a sum of univariate functions of each predictor variable. GAMs do not impose parametric assumptions on the univariate functions, and thus can capture much more complex relationships than GLMs. Moreover, the curse of dimensionality is mitigated since GAMs only require the estimation of univariate functions. However, the assumption that the response variable depends on the predictors through additive effects may still be restrictive in some applications.

The flexibility of GAMs can be further enhanced by incorporating the idea of projection pursuit \cite{kruskal1969toward,friedman1974projection}, which aims to find the most ``interesting'' projection directions of multivariate data. This leads to the development of \textit{generalized additive index models} (GAIMs) \cite{friedman1981projection,roosen1993logistic}, which model the response variable as a sum of univariate functions of linear combinations of the predictor variables. Concretely, a GAIM relates a scalar response \(y\) and a \(d\)-dimensional predictor \(x\) by
\begin{equation}
    g(\mathbb{E}[y|x])=\sum_{j=1}^m f_j(\alpha_j^\top x),\label{eq:gaim}
\end{equation}
where \(g:\mathbb{R}\to\mathbb{R}\) is a known link function, \(m\) is the number of projection indices, and the unknown components to be estimated are the projection indices \(\alpha_1,\dots,\alpha_m\in\mathbb{R}^d\) and the ridge functions \(f_1,\dots,f_m:\mathbb{R}\to\mathbb{R}\). The model \eqref{eq:gaim} with the identity link function is also known as \textit{projection pursuit regression} (PPR) \cite{friedman1981projection}.

GAIMs belong to the class of semiparametric models, where the projection indices are the parametric components and the ridge functions are the nonparametric components, and have the benefits of both parametric and nonparametric models. GAIMs can represent a much larger class of functions than GAMs; it has been proved that the right-hand side of \eqref{eq:gaim}, with proper classes of univariate functions, is a universal approximator for continuous or square integrable multivariate functions as the number of projection indices \(m\) approaches infinity \cite{diaconis1984nonlinear}, though in practice we typically use a small number for \(m\), often smaller than the dimensionality \(d\). Similar to GAMs, GAIMs are free of estimating multivariate regression functions, and thus suffer less from the curse of dimensionality than nonparametric regression. Moreover, the extraction of projection indices in GAIMs provides insights into the underlying structure of the data; as a result, GAIMs can also be preferable to flexible regression methods such as boosting, random forests and neural networks when interpretability is a concern.

While the statistical properties of GAIMs have been extensively studied---including bias and variance analysis for PPR \cite{hall1989projection}, dimensionality-free convergence rates of PPR \cite{chen1991estimation}, and identifiability conditions for GAIMs \cite{yuan2011identifiability}---their computational aspects have received considerably less attention. When initially proposed, PPR was implemented by a forward stage-wise procedure with optional backfitting \cite{friedman1981projection}, and this framework remains the dominant computational approach for GAIMs. In this framework, the estimation of each component \(f_j(\alpha_j^\top x)\) is done sequentially; within each component, \(\alpha_j\) and \(f_j\) are updated alternatively until convergence by fitting the residuals; optionally, previous components can be reestimated after estimating the current component. This procedure has been extended from PPR to GAIMs by constructing residuals via linearizing the link function \cite{roosen1993logistic}. Estimation algorithms for GAIMs outside this framework include \cite{chen2016generalized}, motivated by the need to impose shape constraints on the ridge functions; nevertheless, it relies on the shape restrictions and does not apply to general GAIMs.

Despite its popularity, this framework suffers from several computational drawbacks. First, the sequential estimation of components is highly inefficient, particularly when the number of projection indices \(m\) is large. Furthermore, the estimation error of each component propagates to the other components, sometimes making backfitting a necessity rather than an optional step to obtain good estimation performance, which further increases the computational burden. In a refinement of this procedure \cite{lingjaerde1998generalized}, the algorithm alternates between updating all projection indices and updating all ridge functions; however, the ridge functions are still estimated sequentially. Moreover, in the stage-wise procedure, when estimating each component, the algorithm alternates between computing the optimal projection index given the current ridge function and computing the optimal ridge function given the current projection index via nonparametric smoothing, where the approximate optimality of both the projection index and the ridge function may not be necessary. The algorithm in \cite{ruan2010dimension} applies one-step update to the projection indices for each iteration. However, nonparametric smoothing techniques, which are computationally intensive, remain the major choice for the ridge function step across related algorithms.

In this work, we contribute to the computational aspect of GAIMs by proposing estimation algorithms that overcome the aforementioned drawbacks. First, we apply basis expansion to the ridge functions. This is theoretically grounded in the study of projection estimators in nonparametric regression \cite{tsybakov2009introduction}; in practice, the R package \texttt{mgcv} \cite{wood2011fast,wood2017generalized}, a widely-used implementation for GAMs, uses basis expansion for the estimation of the univariate functions. Basis expansion converts the semiparametric estimation of GAIMs into a parametric estimation problem. Thus, we avoid computationally expensive nonparametric smoothing techniques. Based on this formulation, the estimation task can be viewed as a finite-dimensional optimization problem, which can be solved by gradient descent (GD). GD alternately updates all projection indices and all basis coefficients (thus all ridge functions) with one gradient step, which circumvents all the issues associated with stage-wise procedures. Moreover, with basis expansion, GAIMs (with fixed projection indices) are structurally similar to GLMs. Motivated by the \textit{variational inequality} (VI) estimation for GLMs \cite{juditsky2019signal} and its potential benefits over GD \cite{zhu2025beyond}, we propose a VI-based estimation algorithm for GAIMs. Both GD and VI solve nonconvex optimization problems, and we establish a unified convergence result showing that both algorithms converge to a stationary point at a rate of \(\mathcal{O}(T^{-1/2})\), where \(T\) is the number of iterations. By numerical experiments, we demonstrate the advantages of our proposed methods over the classical stage-wise procedure and the potential benefits of VI over GD.

\section{Proposed Methods}

\subsection{Basis Expansion for Ridge Functions}

We consider the GAIM given in \eqref{eq:gaim}. By basis expansion, we have
\begin{equation}
    f_j(t)=\sum_{k=1}^K \beta_{jk} \phi_k(t),
\end{equation}
where \(\{\phi_k\}_{k=1}^K\) is a set of given basis functions. As a result, the model can be rewritten as
\begin{equation}
    g(\mathbb{E}[y|x])=\sum_{j=1}^m \sum_{k=1}^K \beta_{jk} \phi_k(\alpha_j^\top x)=:f(x;\alpha,\beta).\label{eq:model_basis}
\end{equation}
Given independent and identically distributed (i.i.d.) samples \(\{(x_i,y_i)\}_{i=1}^n\) from the model, the unknown parameters to be estimated are the projection indices and the coefficients of the basis functions,
\[\alpha = (\alpha_1, \dots, \alpha_m) \in \mathbb{R}^{d\times m}, \quad \beta = (\beta_{jk}) \in \mathbb{R}^{m\times K}.\]
With basis expansion, the estimation of the ridge functions is reduced to the estimation of the basis coefficients. Thus, no nonparametric smoothing techniques are needed, and the estimation of GAIMs is converted into a parametric estimation problem.

\subsection{Gradient Descent for GAIMs}

The parametric estimation problem in \eqref{eq:model_basis} can be cast as a finite-dimensional optimization problem. As in the estimation of GLMs, we can often specify a loss function \(\ell(y, \mu)\). For example, if we specify a conditional distribution of \(y|x\), the negative log-likelihood can be used as the loss function. With the loss function specified, we can define the empirical risk
\begin{equation}
    L_n(\alpha, \beta) = \frac{1}{n} \sum_{i=1}^n \ell(y_i, \mu_i),
\end{equation}
where \(\mu_i = g^{-1}(f(x_i; \alpha, \beta))\). Then, the parameters \(\alpha\) and \(\beta\) can be estimated by solving the optimization problem \(\min_{\alpha, \beta} L_n(\alpha, \beta)\), which is known as empirical risk minimization. This optimization problem can be solved by first-order methods such as alternating gradient descent. Specifically, the updates of the parameters at iteration \(t\) are given by
\begin{align}
    \beta^{(t+1)} &= \beta^{(t)} - \gamma_{\beta,t} \nabla_{\beta} L_n\left(\alpha^{(t)}, \beta^{(t)}\right), \label{eq:gd_beta}\\
    \alpha^{(t+1)} &= \Pi_{\mathcal{A}}\left(\alpha^{(t)} - \gamma_{\alpha,t} \nabla_{\alpha} L_n\left(\alpha^{(t)}, \beta^{(t+1)}\right)\right), \label{eq:gd_alpha}
\end{align}
where \(\gamma_{\beta,t}\) and \(\gamma_{\alpha,t}\) are the step sizes for updating \(\beta\) and \(\alpha\). The operator \(\Pi_{\mathcal{A}}\) is the projection onto the constraint set \(\mathcal{A}\) for \(\alpha\). In view of the identifiability conditions for GAIMs \cite{yuan2011identifiability}, we take \(\mathcal{A}=\{\alpha: \|\alpha_j\|_2=1, j=1,\dots,m\}\). We refer to this algorithm as GD.

GD does not suffer from the issues associated with stage-wise procedures discussed in the introduction. All components are handled simultaneously, and for each iteration, only one gradient step is taken for both \(\alpha\) and \(\beta\) instead of solving them to optimality. 

To ensure that GD is a valid algorithm, we need to ensure the differentiability of \(L_n(\alpha, \beta)\) with respect to \(\alpha\) and \(\beta\). This requires the differentiability of the loss function \(\ell\) and the link function \(g\). To take the derivative with respect to \(\alpha\), we also need to ensure the almost-sure differentiability of the basis functions \(\{\phi_k\}_{k=1}^K\).

\subsection{Variational Inequality for GAIMs}

For fixed projection indices \(\alpha\), the model \eqref{eq:model_basis} can be viewed as a GLM with covariates \(\Phi(x;\alpha) = (\phi_k(\alpha_j^\top x))\in\mathbb{R}^{m\times K}\) and parameters \(\beta\in\mathbb{R}^{m\times K}\). This motivates the use of the variational inequality (VI) estimation procedure for GLMs \cite{juditsky2019signal} to update \(\beta\), which solves a VI problem instead of a loss minimization problem and exhibits potential benefits \cite{zhu2025beyond}. Specifically, the VI operator for \(\beta\) is given by \cite{juditsky2019signal}
\begin{equation}
    V_n^{\beta}(\alpha,\beta)=\frac1n\sum_{i=1}^n (\mu_i-y_i)\Phi(x_i;\alpha)\in\mathbb{R}^{m\times K}.
\end{equation}
To fully extend the VI framework from GLMs to GAIMs, a compatible VI operator for \(\alpha\) is required.
Observing that \(\Phi(x;\alpha)\) is the gradient of \(f(x;\alpha,\beta)\) with respect to \(\beta\), we propose to update \(\alpha\) by the gradient of \(f(x;\alpha,\beta)\) with respect to \(\alpha\), which is given by
\begin{equation}
    V_n^{\alpha}(\alpha,\beta)=\frac1n\sum_{i=1}^n (\mu_i-y_i)\nabla_{\alpha} f(x_i;\alpha,\beta)\in\mathbb{R}^{d\times m}.
\end{equation}
Then, we replace the gradients in the GD updates \eqref{eq:gd_beta} and \eqref{eq:gd_alpha} with the VI operators, leading to the updates
\begin{align}
    \beta^{(t+1)} &= \beta^{(t)} - \gamma_{\beta,t} V_n^{\beta}(\alpha^{(t)}, \beta^{(t)}), \\
    \alpha^{(t+1)} &= \Pi_{\mathcal{A}}\left(\alpha^{(t)} - \gamma_{\alpha,t} V_n^{\alpha}(\alpha^{(t)}, \beta^{(t+1)})\right).
\end{align}
We refer to this algorithm as VI.

Compared to GD, VI only requires the almost-sure differentiability of the basis functions \(\{\phi_k\}_{k=1}^K\), but does not require the differentiability of the link function \(g\) and is loss-function agnostic. As a result, VI can accommodate a wider range of link functions.

When we specify a conditional distribution of \(y|x\) and use the negative log-likelihood as the loss function in GD, if the conditional distribution and the link function form a canonical pair, e.g., Gaussian distribution with identity link and Poisson distribution with log link, the updates of GD and VI are equivalent. In other cases, GD and VI may not coincide. For GLMs, some benefits of VI over GD have been observed in \cite{zhu2025beyond}. We provide numerical evidence in Section \ref{sec:vi_gd} that the benefits may remain in the extension to GAIMs.

\section{Theoretical Analysis}

We provide the convergence guarantee for GD and VI by showing they both minimize a common objective \(F_n\) using alternating projected gradient descent. Specifically, let \(F_n=L_n\) for GD. For VI, we define the potential function \(F_n=Q_n\), where
\begin{equation*}
    Q_n(\alpha,\beta) = \frac{1}{n}\sum_{i=1}^n \int_0^{f(x_i;\alpha,\beta)} \left(g^{-1}(s)-y_i\right)ds.
\end{equation*}
By Leibniz's rule, we have \(\nabla_\beta Q_n=V_n^\beta\) and \(\nabla_\alpha Q_n=V_n^\alpha\), so the VI updates are exactly alternating projected gradient descent applied to \(Q_n\). When \(g\) is the canonical link corresponding to the conditional distribution of \(y|x\), \(Q_n\) coincides with \(L_n\) up to an additive constant.

\begin{assumption}\label{ass:converge}
The iterates \(\{(\alpha^{(t)}, \beta^{(t)})\}\) remain in a compact set \(\Omega\subset\mathcal{A}\times\mathbb{R}^{mK}\), on which \(F_n\) is bounded from below. \(\nabla F_n\) is \(L\)-Lipschitz on an open neighborhood of a compact set containing all iterates and blockwise update segments.
The step sizes satisfy \(0<\underline\gamma_\beta\le \gamma_{\beta,t}\le \bar\gamma_\beta<2/L\) and \(0<\underline\gamma_\alpha\le \gamma_{\alpha,t}\le \bar\gamma_\alpha<1/L\).
\end{assumption}

\begin{theorem}\label{thm:converge}
Under Assumption \ref{ass:converge}, every accumulation point \((\bar\alpha,\bar\beta)\) of the iterates generated by GD or VI is a first-order stationary point satisfying
\begin{equation*}
    \nabla_\beta F_n(\bar\alpha,\bar\beta)=0,\quad (I-\bar\alpha_j\bar\alpha_j^\top) \nabla_{\alpha_j}F_n(\bar\alpha,\bar\beta)=0, \, j\in [m].
\end{equation*}
If \(F_n+\iota_{\mathcal A}\) further satisfies the Kurdyka--{\L}ojasiewicz property (e.g., when \(\ell\), \(g\), and \(\{\phi_k\}\) are real analytic), the bounded sequence converges to a stationary point. Moreover, the algorithms achieve an iteration complexity of
\begin{multline*}
    \min_{0\le t<T} \Bigg(\, 
    \sum_{j=1}^m \left\|\left(I-\alpha_j^{(t)}{\alpha_j^{(t)}}^\top\right) \nabla_{\alpha_j}F_n\left(\alpha^{(t)},\beta^{(t)}\right)\right\|_2 +\left\|\nabla_\beta F_n\left(\alpha^{(t)},\beta^{(t)}\right)\right\|_F \Bigg) = \mathcal{O}(T^{-1/2}).
\end{multline*}
\end{theorem}
These claims follow from the PALM framework of \cite{bolte2014proximal}, where the constraint is encoded by the proper, lower-semicontinuous (nonconvex) indicator \(\iota_{\mathcal A}\) with proximal operator given by sphere projection; the Kurdyka--{\L}ojasiewicz convergence machinery therein extends classical block-coordinate descent \cite{tseng2001convergence}, and the \(\mathcal{O}(T^{-1/2})\) rate follows from the nonconvex block-coordinate analysis of \cite{hong2017iteration}.

\section{Numerical Experiments}
In this section, we numerically evaluate the proposed GD and VI algorithms. Across all experiments, the covariates are drawn uniformly from the \(d\)-dimensional unit ball. We set \(K=3\), and use the Legendre polynomials of degree~\(1\)~to~\(3\), subtracted by their values at zero to ensure identifiability, as the basis functions \(\{\phi_k\}_{k=1}^K\). The true projection indices \(\alpha^*\) and the basis coefficients \(\beta^*\) for the true ridge functions are predetermined and fixed for each setting. For the proposed GD and VI algorithms, we initialize the projection indices \(\alpha\) with the first \(m\) standard basis vectors in \(\mathbb{R}^d\), and initialize all basis coefficients \(\beta\) with zeros. Compared to the true parameters to be specified in the following subsections, this initialization is not intended to be close. We consider two evaluation metrics, the projection index estimation error \(\|\hat{\alpha} - \alpha^*\|^2 / \|\alpha^*\|^2\) and the function estimation error \(\mathbb{E}[(\hat{f} - f^*)^2]\), where the expectation is with respect to the distribution of the covariates, approximated using \(10000\) independently drawn test samples. Here, \(f^*\) is the true regression function as in~\eqref{eq:gaim}~and~\eqref{eq:model_basis}, \(\hat{\alpha}\) and \(\hat{\beta}\) are the estimated parameters, and \(\hat{f}(x)=\sum_{j=1}^m \sum_{k=1}^K \hat{\beta}_{jk} \phi_k(\hat{\alpha}_j^\top x)\)~is the estimated regression function. Since the projection indices can be permuted and sign-flipped without changing the model, the projection index estimation error is computed by minimizing over all possible permutations and sign flips of the columns of \(\hat{\alpha}\). Results are averaged over \(1000\) independent trials.

\subsection{Validity of Proposed Methods} \label{sec:vi_gd}

\begin{figure}[htbp]
    \centering
    \begin{subfigure}[t]{0.4\textwidth}
        \centering
        \includegraphics[width=\linewidth]{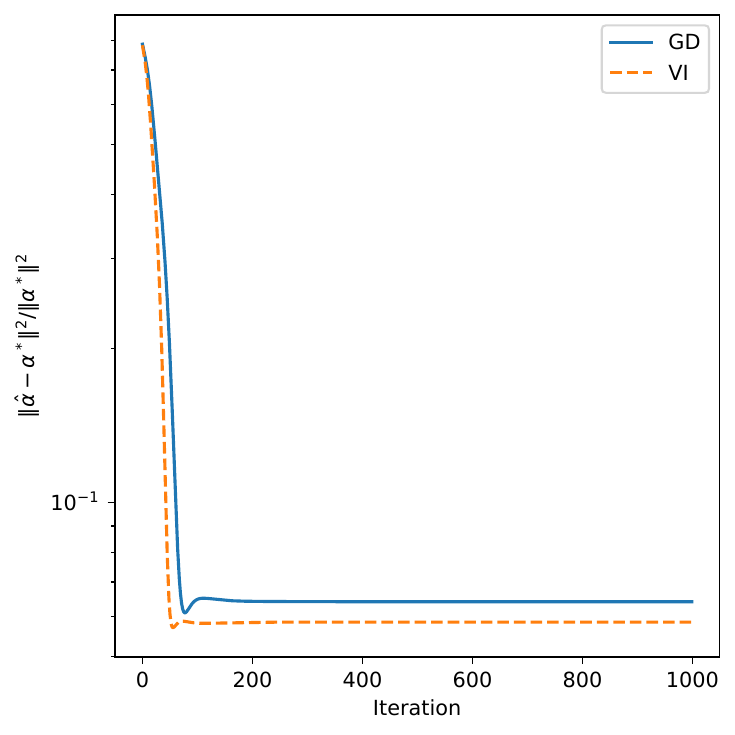}
        \caption{Projection index estimation error}
    \end{subfigure}
    \qquad\quad
    \begin{subfigure}[t]{0.4\textwidth}
        \centering
        \includegraphics[width=\linewidth]{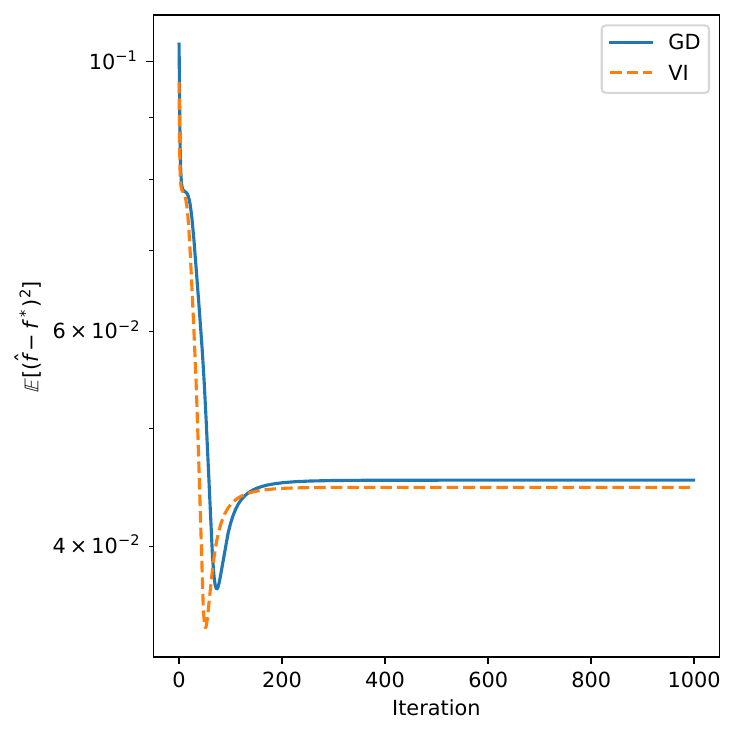}
        \caption{Function estimation error}
    \end{subfigure}
    \caption{Estimation errors of GD and VI for one trial of Poisson additive index models with the inverse softplus link and sample size \(n=2000\).}
    \label{fig:vi_gd_error_plots}
\end{figure}

We verify the validity of the proposed methods, namely GD and VI, by numerical experiments on synthetic data. We keep the dimensionality \(d=4\), the number of projection indices \(m=2\), and vary the sample size \(n \in \{400, 2000, 10000\}\). For this setting, the true parameters are set to \(\alpha_1^* = (1/2, 1/2, 1/2, 1/2)^\top\), \(\alpha_2^* = (1/\sqrt{2}, 1/\sqrt{2}, 0, 0)^\top\), \(\beta_1^* = (1/\sqrt{3}, 1/\sqrt{3}, 1/\sqrt{3})^\top\), and \(\beta_2^* = (1/\sqrt{2}, -1/2, 1/2)^\top\). The responses are generated from Poisson distributions with means given by the true model. We consider two link functions: the canonical log link, for which GD and VI are equivalent, and the inverse softplus link, for which they are not. We run both GD and VI for \(1500\) iterations for the inverse softplus link with \(n=400\) and \(1000\) iterations for other settings. For both algorithms, we use constant step sizes \(\gamma_{\alpha,t}=\gamma_{\beta,t}=1.0\) for the log link, and \(\gamma_{\alpha,t}=\gamma_{\beta,t}=4.0\) for the inverse softplus link. We report the mean metrics with standard deviations in parentheses. Figure~\ref{fig:vi_gd_error_plots} visualizes the estimation errors against the number of iterations for one trial of \(n=2000\) with the inverse softplus link.

The results are summarized in Table~\ref{tab:validity}. As the sample size \(n\) increases, both the projection index estimation error and the function estimation error decrease for both GD and VI under both link functions, confirming the statistical consistency of the proposed methods. For the inverse softplus link, VI outperforms GD in terms of estimation errors of both the projection indices and the function by a small margin across all sample sizes, which indicates that VI may have potential benefits over GD when the link function is not canonical.

\begin{table}[htbp]
\renewcommand{\arraystretch}{1.3}
\caption{Estimation errors of GD and VI for Poisson additive index models under different sample sizes and link functions.}
\label{tab:validity}
\centering
\resizebox{0.8\textwidth}{!}{%
\begin{tabular}{cccccc}
\toprule
\multicolumn{2}{c}{Setting} & \multicolumn{2}{c}{\(\|\hat{\alpha} - \alpha^*\|^2 / \|\alpha^*\|^2\)} & \multicolumn{2}{c}{\(\mathbb{E}[(\hat{f} - f^*)^2]\)} \\
Inverse Link & \(n\) & GD & VI & GD & VI \\
\midrule
\multirow{3}{*}{Exponential} & 400 & \(0.423\) (\(0.431\)) & \(0.423\) (\(0.431\)) & \(0.0594\) (\(0.0288\)) & \(0.0594\) (\(0.0288\)) \\
& 2000 & \(0.051\) (\(0.167\)) & \(0.051\) (\(0.167\)) & \(0.0083\) (\(0.0088\)) & \(0.0083\) (\(0.0088\)) \\
& 10000 & \(0.004\) (\(0.003\)) & \(0.004\) (\(0.003\)) & \(0.0012\) (\(0.0006\)) & \(0.0012\) (\(0.0006\)) \\
\midrule
\multirow{3}{*}{Softplus} & 400 & \(0.726\) (\(0.415\)) & \(\mathbf{0.708}\) (\(0.401\)) & \(0.1744\) (\(0.0648\)) & \(\mathbf{0.1713}\) (\(0.0626\)) \\
& 2000 & \(0.333\) (\(0.464\)) & \(\mathbf{0.315}\) (\(0.448\)) & \(0.0341\) (\(0.0220\)) & \(\mathbf{0.0336}\) (\(0.0218\)) \\
& 10000 & \(0.029\) (\(0.137\)) & \(\mathbf{0.025}\) (\(0.118\)) & \(0.0044\) (\(0.0062\)) & \(\mathbf{0.0043}\) (\(0.0056\)) \\
\bottomrule
\end{tabular}%
}
\end{table}

\subsection{Comparison with Stage-wise Procedures}

We compare our proposed methods with the stage-wise procedure with backfitting \cite{friedman1981projection} implemented in the Python package \texttt{projection-pursuit} \cite{projection-pursuit} (this algorithm is also named as projection pursuit regression, same as the model name). Since this package only supports the identity link function, we restrict the comparison to the Gaussian setting with identity link. In this setting, GD and VI are equivalent.

\begin{figure}[t]
    \centering
    \begin{subfigure}[t]{0.4\textwidth}
        \centering
        \includegraphics[width=\linewidth]{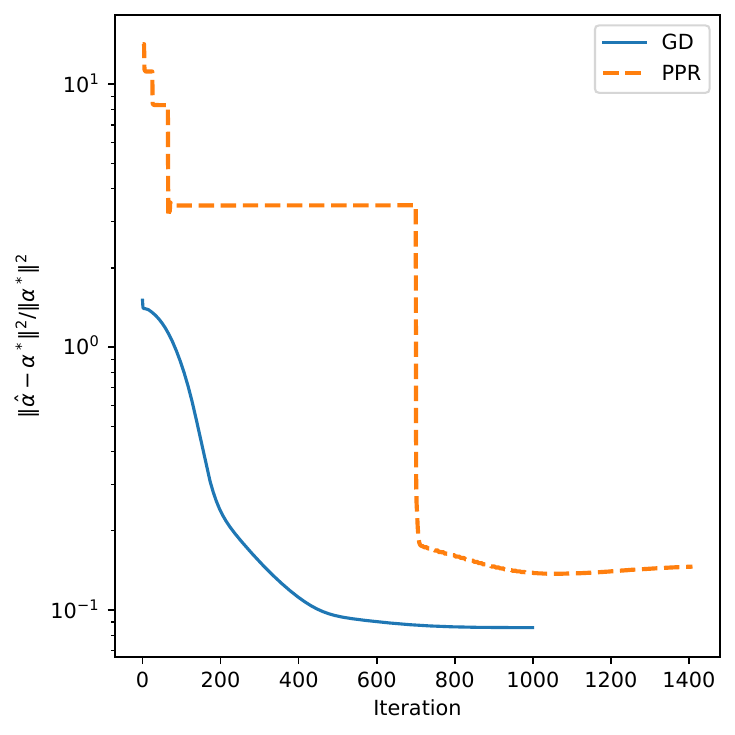}
        \caption{Projection index estimation error}
    \end{subfigure}
    \qquad\quad
    \begin{subfigure}[t]{0.4\textwidth}
        \centering
        \includegraphics[width=\linewidth]{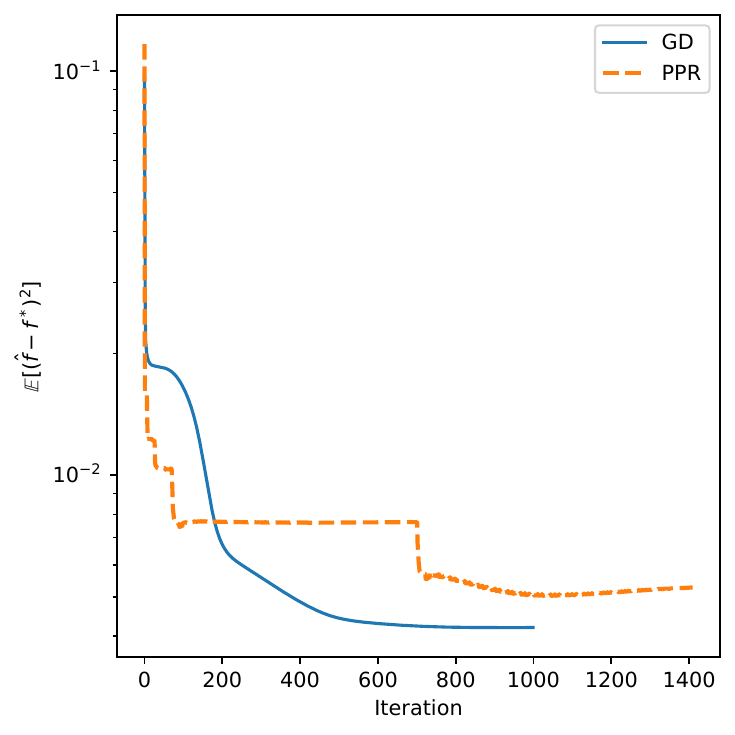}
        \caption{Function estimation error}
    \end{subfigure}
    \caption{Estimation errors of GD and PPR for one trial of Gaussian additive index models with \((d,m)=(20,5)\) and sample size \(n=2000\).}
    \label{fig:ppr_error_plots}
\end{figure}

We keep the sample size fixed at \(n=2000\) and vary the dimensionality \(d\) and the number of projection indices \(m\), considering \((d,m)=(4,2), (20,5), (50,10)\).  For each setting, \(\alpha^*\) is constructed as a block-sparse orthogonal matrix where each column has equal non-zero entries in disjoint blocks of size \(d/m\), and \(\beta^*\) is generated with structurally unique sign patterns and row-dependent decay across the \(m\) components, normalized such that \(\|\beta_j^*\|_2 = 1\). The responses are generated from Gaussian distributions with means given by the true model and variances \(0.125, 0.0625, 0.05\) for \((d,m)=(4,2), (20,5), (50,10)\) respectively. Here, the variances are proportional to \(m/d\) to maintain a moderate and nearly identical signal-to-noise ratio across different settings. For \((d,m)=(4,2)\), we run GD for \(200\) iterations with step sizes \(\gamma_{\alpha,t}=\gamma_{\beta,t}=0.3\); for \((d,m)=(20,5)\), we run GD for \(1000\) iterations with step sizes \(\gamma_{\alpha,t}=\gamma_{\beta,t}=0.5\); for \((d,m)=(50,10)\), we run GD for \(1000\) iterations with step sizes \(\gamma_{\alpha,t}=\gamma_{\beta,t}=0.2\). For PPR, we use cubic polynomial fitting for the ridge function step so that PPR has the same expressive power for ridge functions as GD. It is not straightforward to control the number of iterations for PPR; instead, we set reasonable values for the maximum numbers of iterations for the component-wise updates and the backfitting and the tolerance parameters for convergence. Figure~\ref{fig:ppr_error_plots} visualizes the estimation errors against the number of iterations for one trial of \((d,m)=(20,5)\). Here, the number of iterations for PPR increases by one whenever the ridge function of any component is updated. Due to the vast difference in the nature of GD and PPR, the numbers of iterations and the trends of estimation errors against iterations may not be directly comparable. Therefore, we choose to report the final estimation errors and the total running times for both algorithms in Table~\ref{tab:ppr_comparison}.

The results are summarized in Table~\ref{tab:ppr_comparison}. It can be observed that with a moderate sample size \(n=2000\), GD consistently outperforms the stage-wise procedure in terms of running time and estimation errors, except for running time when \((d,m)=(4,2)\). Due to the sequential nature and use of second-order methods for updating the projection indices, the running time of the stage-wise procedure increases significantly as \(d\) and \(m\) grow. In contrast, the running time of GD exhibits a much more benign scaling with \(d\) and \(m\), allowing it to achieve competitive estimation errors with significantly shorter running times as \(d\) and \(m\) increase.

\begin{table}[t]
\renewcommand{\arraystretch}{1.3}
\caption{Comparison between GD and PPR for additive index models.}
\label{tab:ppr_comparison}
\centering
\resizebox{0.8\textwidth}{!}{%
\begin{tabular}{cccccccc}
\toprule
\multicolumn{2}{c}{Setting} & \multicolumn{2}{c}{Running Time (s)} & \multicolumn{2}{c}{\(\|\hat{\alpha} - \alpha^*\|^2 / \|\alpha^*\|^2\)} & \multicolumn{2}{c}{\(\mathbb{E}[(\hat{f} - f^*)^2]\)} \\
\(d\) & \(m\) & GD & PPR & GD & PPR & GD & PPR \\
\midrule
4 & 2 & \(0.12\) (\(0.02\)) & \(\mathbf{0.01}\) (\(0.00\)) & \(\mathbf{0.0025}\) (\(0.0258\)) & \(0.0108\) (\(0.0038\)) & \(\mathbf{0.0008}\) (\(0.0021\)) & \(0.0021\) (\(0.0004\)) \\
20 & 5 & \(\mathbf{0.46}\) (\(0.07\)) & \(0.79\) (\(1.64\)) & \(\mathbf{0.1941}\) (\(0.1667\)) & \(0.2388\) (\(0.1750\)) & \(\mathbf{0.0061}\) (\(0.0026\)) & \(0.0067\) (\(0.0023\)) \\
50 & 10 & \(\mathbf{1.25}\) (\(0.18\)) & \(4.34\) (\(1.74\)) & \(\mathbf{1.3048}\) (\(0.0607\)) & \(1.3903\) (\(0.0519\)) & \(\mathbf{0.0119}\) (\(0.0015\)) & \(0.0594\) (\(0.0045\)) \\
\bottomrule
\end{tabular}%
}
\end{table}

\section{Discussion}

In this paper, we addressed the computational challenges associated with the estimation of generalized additive index models (GAIMs). By applying basis expansion to the nonparametric ridge functions, we transformed the semiparametric estimation of GAIMs into a finite-dimensional parametric optimization problem, bypassing the need for expensive nonparametric smoothing techniques required by classical stage-wise procedures. We proposed two algorithms, Gradient Descent (GD) and a Variational Inequality (VI) approach, to solve this problem simultaneously across all components. Our theoretical analysis provided a unified convergence guarantee for both methods, establishing a convergence rate of \(\mathcal{O}(T^{-1/2})\) to a stationary point. Empirically, our GD method demonstrated computational and statistical advantages over the classical stage-wise procedure with backfitting, especially as the dimensionality and the number of indices grow, while our VI method exhibited potential benefits over GD for non-canonical link functions.

There are several promising directions for future research. First, it is of statistical interest to understand the recovery guarantees and sample complexities of estimating the projection indices and the ridge functions for the proposed methods. Second, it is practically relevant to characterize the impact of the choice of basis functions. Finally, both of our proposed methods can be straightforwardly adapted to the stochastic setting, where the updates are based on mini-batches of samples instead of the full dataset. This is particularly appealing for large datasets and can potentially further improve the computational efficiency.

\appendix

\section{Proof of Theorem \ref{thm:converge}}

For notational simplicity, write \(F^{(t)}:=F_n(\alpha^{(t)},\beta^{(t)})\),
and define the increments
\[
    d_\beta^{(t)}
    :=
    \beta^{(t+1)}-\beta^{(t)},
    \qquad
    d_\alpha^{(t)}
    :=
    \alpha^{(t+1)}-\alpha^{(t)} .
\]

First, we prove sufficient descent. Since \(\nabla F_n\) is \(L\)-Lipschitz,
the descent lemma gives
\begin{align*}
    F_n(\alpha^{(t)},\beta^{(t+1)})
    &\le
    F_n(\alpha^{(t)},\beta^{(t)})
    +
    \left\langle
    \nabla_\beta F_n(\alpha^{(t)},\beta^{(t)}),
    d_\beta^{(t)}
    \right\rangle+
    \frac{L}{2}\|d_\beta^{(t)}\|_F^2 .
\end{align*}
Using the \(\beta\)-update
\(d_\beta^{(t)}
    =
    -
    \gamma_{\beta,t}
    \nabla_\beta F_n(\alpha^{(t)},\beta^{(t)})\),
we obtain
\begin{equation}
    F_n(\alpha^{(t)},\beta^{(t+1)})
    \le
    F^{(t)}
    -
    \left(
    \frac1{\gamma_{\beta,t}}-\frac L2
    \right)
    \|d_\beta^{(t)}\|_F^2 .
    \label{eq:proof_beta_descent}
\end{equation}

Next, define
\[
    G_\alpha^{(t)}
    :=
    \nabla_\alpha F_n(\alpha^{(t)},\beta^{(t+1)}),
    \qquad
    z^{(t)}
    :=
    \alpha^{(t)}-\gamma_{\alpha,t}G_\alpha^{(t)} .
\]
Since
\(
    \alpha^{(t+1)}
    =
    \Pi_{\mathcal A}(z^{(t)})
\)
and \(\alpha^{(t)}\in\mathcal A\), the projection property gives
\(
    \|\alpha^{(t+1)}-z^{(t)}\|_F^2
    \le
    \|\alpha^{(t)}-z^{(t)}\|_F^2\).
Substituting \(z^{(t)}=\alpha^{(t)}-\gamma_{\alpha,t}G_\alpha^{(t)}\)
and \(d_\alpha^{(t)}=\alpha^{(t+1)}-\alpha^{(t)}\) yields
\[
    \|d_\alpha^{(t)}
    +
    \gamma_{\alpha,t}G_\alpha^{(t)}\|_F^2
    \le
    \gamma_{\alpha,t}^2\|G_\alpha^{(t)}\|_F^2 .
\]
After expanding both sides, we get
\begin{equation*}
    \left\langle
    G_\alpha^{(t)},d_\alpha^{(t)}
    \right\rangle
    \le
    -
    \frac1{2\gamma_{\alpha,t}}
    \|d_\alpha^{(t)}\|_F^2 .
    \label{eq:projection_inner_product}
\end{equation*}
Applying the descent lemma to the \(\alpha\)-block gives
\begin{align}
    F^{(t+1)}
    &=
    F_n(\alpha^{(t+1)},\beta^{(t+1)})\nonumber\\
    &\le
    F_n(\alpha^{(t)},\beta^{(t+1)})
    +
    \left\langle
    G_\alpha^{(t)},d_\alpha^{(t)}
    \right\rangle
    +
    \frac L2\|d_\alpha^{(t)}\|_F^2
    \nonumber\\
    &\le
    F_n(\alpha^{(t)},\beta^{(t+1)})
    -
    \left(
    \frac1{2\gamma_{\alpha,t}}-\frac L2
    \right)
    \|d_\alpha^{(t)}\|_F^2 .
    \label{eq:proof_alpha_descent}
\end{align}
Combining \eqref{eq:proof_beta_descent} and
\eqref{eq:proof_alpha_descent}, we obtain
\begin{equation}
    F^{(t+1)}
    \le
    F^{(t)}
    -
    c_\beta
    \|d_\beta^{(t)}\|_F^2
    -
    c_\alpha
    \|d_\alpha^{(t)}\|_F^2 ,
    \label{eq:sufficient_descent}
\end{equation}
where
\(
    c_\beta
    :=
    \frac1{\bar\gamma_\beta}-\frac L2>0\),
     \(c_\alpha
    :=
    \frac1{2\bar\gamma_\alpha}-\frac L2>0\).
Let \(c_{\min}:=\min\{c_\beta,c_\alpha\}\).
Since \(F_n\) is bounded from below on \(\Omega\), summing
\eqref{eq:sufficient_descent} from \(t=0\) to \(T-1\) gives
\begin{equation}
    \sum_{t=0}^{T-1}
    \left(
    \|d_\beta^{(t)}\|_F^2
    +
    \|d_\alpha^{(t)}\|_F^2
    \right)
    \le
    \frac{
    F^{(0)}-\inf_{(\alpha,\beta)\in\Omega}F_n(\alpha,\beta)
    }{c_{\min}} .
    \label{eq:summed_steps}
\end{equation}
Letting \(T\to\infty\), we have
\(
    \sum_{t=0}^{\infty}
    (
    \|d_\beta^{(t)}\|_F^2
    +
    \|d_\alpha^{(t)}\|_F^2
    )
    <\infty\).
Therefore,
\begin{equation}
    \|d_\beta^{(t)}\|_F\to0,
    \qquad
    \|d_\alpha^{(t)}\|_F\to0 .
    \label{eq:successive_diff_zero}
\end{equation}

We now prove stationarity of every accumulation point. 
Using \(\gamma_{\beta,t}\ge \underline\gamma_\beta\) and
\eqref{eq:successive_diff_zero}, we obtain
\begin{equation}
    \|\nabla_\beta F_n(\alpha^{(t)},\beta^{(t)})\|_F
    \le
    \frac1{\underline\gamma_\beta}
    \|d_\beta^{(t)}\|_F
    \to0 .
    \label{eq:beta_grad_zero}
\end{equation}
For the \(\alpha\)-block, recall that
\(\mathcal A\) is the product of unit spheres. The projection step implies,
componentwise, that
\[
    \alpha_j^{(t+1)}
    =
    \Pi_{\mathbb S^{d-1}}
    \left(
    \alpha_j^{(t)}
    -
    \gamma_{\alpha,t}
    G_{\alpha,j}^{(t)}
    \right),
    \qquad j=1,\ldots,m .
\]
Hence
\(
    G_{\alpha,j}^{(t)}
    +
    \frac1{\gamma_{\alpha,t}}
    d_{\alpha,j}^{(t)}
\)
is normal to the sphere at \(\alpha_j^{(t+1)}\). Equivalently,
\begin{equation*}
    \left(
    I-\alpha_j^{(t+1)}{\alpha_j^{(t+1)}}^\top
    \right)
    \left(
    G_{\alpha,j}^{(t)}
    +
    \frac1{\gamma_{\alpha,t}}
    d_{\alpha,j}^{(t)}
    \right)
    =
    0 .
    \label{eq:normal_condition_alpha}
\end{equation*}
Therefore,
\begin{align*}
    &\left\|
    \left(
    I-\alpha_j^{(t+1)}{\alpha_j^{(t+1)}}^\top
    \right)
    \nabla_{\alpha_j}F_n(\alpha^{(t+1)},\beta^{(t+1)})
    \right\|_2
    \nonumber\\
    \le\ &
    \left\|
    \nabla_{\alpha_j}F_n(\alpha^{(t+1)},\beta^{(t+1)})
    -
    \nabla_{\alpha_j}F_n(\alpha^{(t)},\beta^{(t+1)})
    \right\|_2+
    \frac1{\gamma_{\alpha,t}}
    \|d_{\alpha,j}^{(t)}\|_2 .
\end{align*}
Summing over \(j\) in squared norm and using the \(L\)-Lipschitz continuity
of \(\nabla F_n\), we get
\begin{align}
    &\left(
    \sum_{j=1}^m
    \left\|
    \left(
    I-\alpha_j^{(t+1)}{\alpha_j^{(t+1)}}^\top
    \right)
    \nabla_{\alpha_j}F_n(\alpha^{(t+1)},\beta^{(t+1)})
    \right\|_2^2
    \right)^{1/2}\le
    \left(
    L+\frac1{\underline\gamma_\alpha}
    \right)
    \|d_\alpha^{(t)}\|_F .
    \label{eq:alpha_grad_zero}
\end{align}
Thus the projected \(\alpha\)-gradient also converges to zero along the
sequence shifted by one step.

Let \((\bar\alpha,\bar\beta)\) be an accumulation point of
\(\{(\alpha^{(t)},\beta^{(t)})\}\). Then there exists a subsequence
\(t_r\to\infty\) such that
\((\alpha^{(t_r)},\beta^{(t_r)})
    \to
    (\bar\alpha,\bar\beta)\).
By \eqref{eq:successive_diff_zero}, we also have
\(
    (\alpha^{(t_r+1)},\beta^{(t_r+1)})
    \to
    (\bar\alpha,\bar\beta).
\)
Taking limits in \eqref{eq:beta_grad_zero} and
\eqref{eq:alpha_grad_zero}, and using continuity of \(\nabla F_n\), yields
\[
    \nabla_\beta F_n(\bar\alpha,\bar\beta)=0,
\]
and
\[
    \left(
    I-\bar\alpha_j\bar\alpha_j^\top
    \right)
    \nabla_{\alpha_j}F_n(\bar\alpha,\bar\beta)=0,
    \qquad j=1,\ldots,m .
\]
Therefore every accumulation point is a first-order stationary point of
the constrained problem.

We next discuss convergence of the whole sequence under the
Kurdyka--{\L}ojasiewicz property. Let
\[
    \Psi(\alpha,\beta)
    :=
    F_n(\alpha,\beta)+\iota_{\mathcal A}(\alpha).
\]
The sufficient descent inequality \eqref{eq:sufficient_descent} gives the
standard sufficient-decrease condition for PALM:
\[
    \Psi(\alpha^{(t+1)},\beta^{(t+1)})
    \le
    \Psi(\alpha^{(t)},\beta^{(t)})
    -
    c_{\min}
    \left(
    \|d_\beta^{(t)}\|_F^2
    +
    \|d_\alpha^{(t)}\|_F^2
    \right).
\]
Moreover, the optimality condition of the projection step gives a
relative-error bound. Indeed, define
\[
    x^{(t)}:=(\alpha^{(t)},\beta^{(t)}),
    \qquad
    d^{(t)}:=x^{(t+1)}-x^{(t)} .
\]
The \(\beta\)-component satisfies
\begin{align*}
    \|\nabla_\beta F_n(x^{(t+1)})\|_F
    &\le
    \|\nabla_\beta F_n(x^{(t)})\|_F
    +
    L\|d^{(t)}\|_F\le
    \left(
    \frac1{\underline\gamma_\beta}+L
    \right)
    \|d^{(t)}\|_F .
\end{align*}
For the \(\alpha\)-component, the projection optimality condition implies
that there exists
\(n_\alpha^{(t+1)}\in N_{\mathcal A}(\alpha^{(t+1)})\) such that
\(
    n_\alpha^{(t+1)}
    =
    -
    G_\alpha^{(t)}
    -
    \frac1{\gamma_{\alpha,t}}d_\alpha^{(t)}\).
Hence
\begin{align*}
    &\nabla_\alpha F_n(x^{(t+1)})+n_\alpha^{(t+1)}
    =
    \nabla_\alpha F_n(\alpha^{(t+1)},\beta^{(t+1)})
    -
    \nabla_\alpha F_n(\alpha^{(t)},\beta^{(t+1)})
    -
    \frac1{\gamma_{\alpha,t}}d_\alpha^{(t)} ,
\end{align*}
and therefore
\[
    \|\nabla_\alpha F_n(x^{(t+1)})+n_\alpha^{(t+1)}\|_F
    \le
    \left(
    L+\frac1{\underline\gamma_\alpha}
    \right)
    \|d^{(t)}\|_F .
\]
Consequently, there exists a constant \(C_{\rm rel}>0\) such that
\[
    \operatorname{dist}
    \left(
    0,\partial\Psi(x^{(t+1)})
    \right)
    \le
    C_{\rm rel}\|x^{(t+1)}-x^{(t)}\|_F .
\]
Thus the iterates satisfy the sufficient-decrease and relative-error
conditions required by the PALM convergence theorem. If
\(\Psi=F_n+\iota_{\mathcal A}\) satisfies the Kurdyka--{\L}ojasiewicz
property, the bounded sequence has finite length,
\[
    \sum_{t=0}^{\infty}
    \|x^{(t+1)}-x^{(t)}\|_F<\infty,
\]
and hence converges to a single stationary point.

It remains to prove the stated iteration complexity. Define the
stationarity residual
\begin{align*}
    R_t
    :=
    &\ \|\nabla_\beta F_n(\alpha^{(t)},\beta^{(t)})\|_F+
    \sum_{j=1}^m
    \left\|
    \left(
    I-\alpha_j^{(t)}{\alpha_j^{(t)}}^\top
    \right)
    \nabla_{\alpha_j}F_n(\alpha^{(t)},\beta^{(t)})
    \right\|_2 .
\end{align*}
For the \(\beta\)-part, the update gives
\begin{equation}
    \|\nabla_\beta F_n(\alpha^{(t)},\beta^{(t)})\|_F
    \le
    \frac1{\underline\gamma_\beta}
    \|d_\beta^{(t)}\|_F .
    \label{eq:complexity_beta}
\end{equation}
For the \(\alpha\)-part, apply the projection optimality condition from the
previous iteration. For \(t\ge1\),
\[
    \left(
    I-\alpha_j^{(t)}{\alpha_j^{(t)}}^\top
    \right)
    \left(
    \nabla_{\alpha_j}F_n(\alpha^{(t-1)},\beta^{(t)})
    +
    \frac1{\gamma_{\alpha,t-1}}
    d_{\alpha,j}^{(t-1)}
    \right)
    =
    0 .
\]
Therefore,
\begin{align}
    &\sum_{j=1}^m
    \left\|
    \left(
    I-\alpha_j^{(t)}{\alpha_j^{(t)}}^\top
    \right)
    \nabla_{\alpha_j}F_n(\alpha^{(t)},\beta^{(t)})
    \right\|_2
    \nonumber\\
    \le\ &
    \sum_{j=1}^m
    \left\|
    \nabla_{\alpha_j}F_n(\alpha^{(t)},\beta^{(t)})
    -
    \nabla_{\alpha_j}F_n(\alpha^{(t-1)},\beta^{(t)})
    \right\|_2+
    \frac1{\underline\gamma_\alpha}
    \sum_{j=1}^m
    \|d_{\alpha,j}^{(t-1)}\|_2
    \nonumber\\
    \le\ &
    \sqrt m
    \left(
    L+\frac1{\underline\gamma_\alpha}
    \right)
    \|d_\alpha^{(t-1)}\|_F .
    \label{eq:complexity_alpha}
\end{align}
Combining \eqref{eq:complexity_beta} and \eqref{eq:complexity_alpha},
for every \(t\ge1\),
\[
    R_t
    \le
    C_\beta\|d_\beta^{(t)}\|_F
    +
    C_\alpha\|d_\alpha^{(t-1)}\|_F ,
\]
where
\(
    C_\beta:=\frac1{\underline\gamma_\beta}\),
     \(C_\alpha:=
    \sqrt m
    \left(
    L+\frac1{\underline\gamma_\alpha}
    \right)\).
Thus
\[
    R_t^2
    \le
    2C_\beta^2\|d_\beta^{(t)}\|_F^2
    +
    2C_\alpha^2\|d_\alpha^{(t-1)}\|_F^2 .
\]
Summing from \(t=1\) to \(T-1\) and using \eqref{eq:summed_steps}, we get
\[
    \sum_{t=1}^{T-1}R_t^2
    \le
    \frac{
    2(C_\beta^2+C_\alpha^2)
    \left[
    F^{(0)}
    -
    \inf_{(\alpha,\beta)\in\Omega}F_n(\alpha,\beta)
    \right]
    }{c_{\min}} .
\]
Therefore, for \(T\ge2\),
\[
    \min_{1\le t<T}R_t^2
    \le
    \frac{
    2(C_\beta^2+C_\alpha^2)
    \left[
    F^{(0)}
    -
    \inf_{(\alpha,\beta)\in\Omega}F_n(\alpha,\beta)
    \right]
    }{
    c_{\min}(T-1)
    } .
\]
Taking square roots gives
\(
    \min_{1\le t<T}R_t
    =
    \mathcal O(T^{-1/2})
\). This completes the proof.

\bibliographystyle{unsrtnat}
\bibliography{ref}

\end{document}